\documentclass[aps,prl,twocolumn,showpacs,preprintnumbers,amsmath,amssymb,superscriptaddress]{revtex4}

\usepackage{graphicx}
\usepackage{dcolumn}
\usepackage{bm}

\begin{document}


\title{Coincidence measurement of residues and light particles in the reaction $^{56}$Fe+p\\ at 1 GeV per nucleon with SPALADIN}

\author{E.~Le Gentil}
\affiliation{DAPNIA/SPhN, CEA/Saclay, France}
\author{T.~Aumann}
\affiliation{GSI, Darmstadt, Germany}
\author{C.O.~Bacri }
\affiliation{IPN, CNRS-IN2P3, Orsay, France}
\author{J.~Benlliure}
\affiliation{University of Santiago de Compostela, Spain}
\author{S.~Bianchin}
\affiliation{GSI, Darmstadt, Germany}
\author{M.~B\"ohmer}
\affiliation{TU M\"unchen, Garching, Germany}
\author{A.~Boudard}
\affiliation{DAPNIA/SPhN, CEA/Saclay, France}
\author{J.~Brzychczyk}
\affiliation{Jagellonian University, Krakow, Poland}
\author{E.~Casarejos}
\affiliation{University of Santiago de Compostela, Spain}
\author{M.~Combet}
\affiliation{DAPNIA/SPhN, CEA/Saclay, France}
\author{L.~Donadille}
\affiliation{DAPNIA/SPhN, CEA/Saclay, France}
\author{J.E.~Ducret}
\affiliation{DAPNIA/SPhN, CEA/Saclay, France}
\author{M.~Fernandez-Ordo\~nez}
\affiliation{University of Santiago de Compostela, Spain}
\author{R.~Gernh\"auser}
\affiliation{TU M\"unchen, Garching, Germany}
\author{H.~Johansson}
\affiliation{GSI, Darmstadt, Germany}
\author{K.~Kezzar}
\affiliation{DAPNIA/SPhN, CEA/Saclay, France}
\author{T.~Kurtukian-Nieto}
\affiliation{University of Santiago de Compostela, Spain}
\author{A.~Lafriakh}
\affiliation{IPN, CNRS-IN2P3, Orsay, France}
\author{F.~Lavaud}
\affiliation{DAPNIA/SPhN, CEA/Saclay, France}
\author{A.~Le F\`evre}
\affiliation{GSI, Darmstadt, Germany}
\author{S.~Leray}
\affiliation{DAPNIA/SPhN, CEA/Saclay, France}
\author{J.~L\"uhning}
\affiliation{GSI, Darmstadt, Germany}
\author{J.~Lukasik}
\affiliation{GSI, Darmstadt, Germany}
\affiliation{H. Niewodniczanski Institute of nuclear physics, Krakow, Poland}
\author{U.~Lynen}
\affiliation{GSI, Darmstadt, Germany}
\author{ W.F.J.~M\"uller}
\affiliation{GSI, Darmstadt, Germany}
\author{P.~Pawlowski}
\affiliation{H. Niewodniczanski Institute of nuclear physics, Krakow, Poland}
\author{S.~Pietri}
\affiliation{DAPNIA/SPhN, CEA/Saclay, France}
\author{F.~Rejmund}
\affiliation{GANIL, CEA \& IN2P3, Caen, France}
\author{C.~Schwarz}
\affiliation{GSI, Darmstadt, Germany}
\author{C. Sfienti}
\affiliation{GSI, Darmstadt, Germany}
\author{H.~Simon}
\affiliation{GSI, Darmstadt, Germany}
\author{W.~Trautmann}
\affiliation{GSI, Darmstadt, Germany}
\author{C.~Volant}
\affiliation{DAPNIA/SPhN, CEA/Saclay, France}
\author{O.~Yordanov}
\affiliation{GSI, Darmstadt, Germany}

\date{\today}

\begin{abstract}
The spallation of $^{56}$Fe in collisions with hydrogen at 1 A GeV has been studied in inverse kinematics with the large-aperture setup SPALADIN at GSI. Coincidences of residues with low-center-of-mass kinetic energy light particles and fragments have been measured allowing the decomposition of the total reaction cross-section into the different possible de-excitation channels. Detailed information on the evolution of these de-excitation channels with excitation energy has also been obtained. The comparison of the data with predictions of several de-excitation models coupled to the INCL4 intra-nuclear cascade model shows that only GEMINI can reasonably account for the bulk of collected results, indicating that in a light system with no compression and little angular momentum, multifragmentation might not be necessary to explain the data.
\end{abstract}

\maketitle

Spallation reactions play an important role in many domains ranging from astrophysics to intense neutron sources. Proton-induced reactions are also a way to study the de-excitation mechanism of a nucleus in a single hot source, and with less dynamical effects than in nucleus-nucleus collisions. They are often described as a 2-step model, with an intra-nuclear cascade (INC) phase followed by a de-excitation phase. Inclusive data on light particles emitted in the spallation process and, more recently, data on spallation residues, helped considerably in improving the models~\cite{HIN05}. However, these are not sufficient to provide a real insight into the reaction mechanism and the discrepancies observed between data and codes cannot be interpreted unambiguously with inclusive data. This is in particular due to the fact that the final observables are often both influenced by the cascade phase (especially by the remnant excitation energy) and by the de-excitation phase. A few more exclusive measurements exist but are generally limited to the study of the most violent collisions representing a small part of the total reaction cross-section (for a review see e.g.~\cite{VIO06}).

The need for a better understanding of spallation reactions motivated the design of the SPALADIN setup at GSI, which aims at measuring in inverse kinematics and {\it in coincidence} all the spallation products with a low center-of-mass (c.m.) kinetic energy, from neutrons to heavy residues. The restriction to low c.m. energies, in fact due to geometrical acceptance limitations, largely favors the detection of particles from the de-excitation rather than the cascade phase. This allows to use the particle multiplicities as an indication of the excitation energy ($E^{\star}$) at the end of the cascade stage.

The SPALADIN setup, partially described in~\cite{LEG06}, is based on the inverse kinematics technique where the ion beam is projected onto a liquid hydrogen target. The use of the large acceptance dipole magnet ALADIN permits to select the particles with a low c.m. kinetic energy. Among other detectors, the setup comprises the large area neutron detector LAND, which provides neutron multiplicities, a time-of-flight wall and the multitrack and multiple-sampling time projection chamber (TPC) MUSIC-IV. The TPC enables a good charge identification for Z down to protons. The use of flash ADC's permits the simultaneous detection and reconstruction of several tracks. Its global efficiency was calculated to be 44~\% for hydrogen, 78~\% for He, 83~\% for Li, and $>94$~\% for heavier fragments. The acceptance of the setup has been estimated through a GEANT~4 simulation as being around 20~\% for protons, 80~\% for helium and $>$~97~\% for heavier fragments. Our relatively poor efficiency and acceptance for protons is the reason why data on protons are not shown here. Data are obtained after empty target subtraction, once normalized to the number of incident beam particles counted with a thin plastic scintillator also used as a trigger. In the following, they will be compared to the codes filtered by GEANT~4.

The first experiment using this setup was performed on the system Fe+p at 1 GeV per nucleon. In addition to the coincidence data, inclusive element cross sections and recoil velocity distributions were obtained, which will be presented in more details in a forthcoming paper. The element cross sections obtained with SPALADIN agree with previous data from the FRagment Separator (FRS) ~\cite{VIL06,NAP04} (not shown here) except for carbon and nitrogen, which deviate by 25\%.

\begin{figure}
\includegraphics[width=7.3cm]{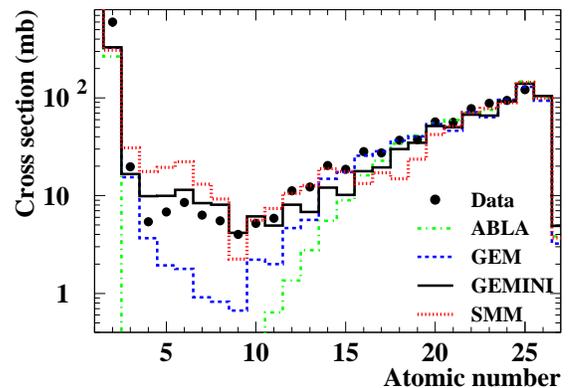}
\caption{\label{secz}
(Color on line) Element production cross sections compared to ABLA, GEMINI, SMM and GEM, associated with INCL4.}
\end{figure}

As was done for the FRS data~\cite{VIL06}, comparisons with several de-excitation models associated with the intra-nuclear cascade INCL4~\cite{BOU02} allows to draw some first conclusions (see Fig.~\ref{secz}). The combination of INCL4 with the de-excitation model ABLA~\cite{JUN98}, which is generally rather successful in predicting data in heavy systems, seriously underpredicts the production of fragments with charges between 3 and 10 (hereafter denoted as Intermediate Mass Fragments or IMF). This can be ascribed to the fact that ABLA is a "classical" de-excitation model in which only the evaporation of neutrons, protons and alpha particles is possible. To account for the emission of IMF, several de-excitation mechanisms can be envisaged: A generalized evaporation, as modeled in GEM~\cite{FUR00} which evaporates particles up to Mg; a binary splitting as described in GEMINI~\cite{CHA88}, which uses the Moretto-type transition state model~\cite{MOR75} for the emission of intermediate mass fragments down to a certain limit, the evaporation of lighter fragments being treated within the Hauser-Feshbach formalism; the opening of multifragmentation channels, as included in SMM~\cite{SMM90}. Default parameters of the models have been used except for GEMINI for which the transition from the Hauser-Feshbach to the Transition State Model is made for Z$\geq$4 as recommended by the author~\cite{CHA05}. We can see that IMF cross-sections are largely under-predicted by GEM while GEMINI better agree with the data and SMM overestimates them. Similar conclusions were drawn also in~\cite{MAS04}. However, from these inclusive data it is not possible to draw more precise conclusions on the actual mechanism.

We cannot exclude that the underprediction of light elements could be due to the intra-nuclear cascade model, INCL4, which, for instance, could predict a too small excitation energy of the de-exciting nucleus. However, it was shown in~\cite{VIL06} that a cascade model leading to higher excitation energies would certainly help populating the IMF region but at the detriment of the high Z part of the spectrum. It was also noticed that excitation energies given by INCL4 at the end of the cascade are remarkably close to those given by the ISABEL code~\cite{YAR79}, which uses a very different modeling of the cascade.

\begin{figure}
\includegraphics[height=8.5cm, angle=-90]{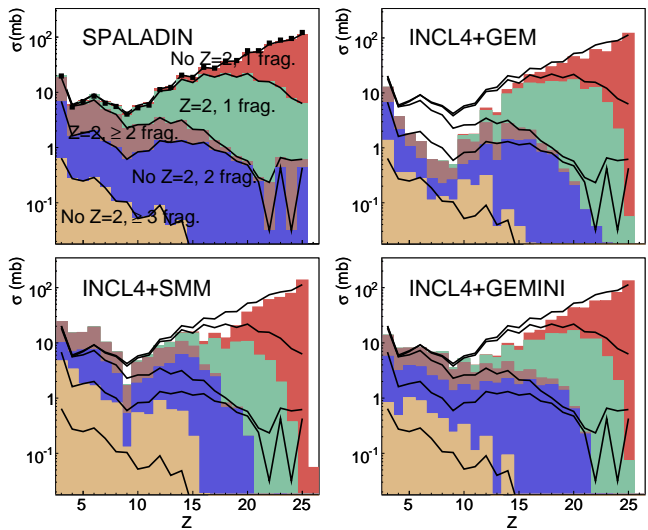}
\caption{\label{contrib} (Color online) Contributions of the different channels to the element cross section, measured (upper-left panel) and calculated with GEM, SMM and GEMINI, associated with INCL4 and filtered through GEANT4. The solid lines delimiting the different regions are taken from the upper-left panel and shown on the other panels. Fragment means Z$\geq$3.}
\end{figure}

With SPALADIN, it is possible to decompose the element production cross section into the different reaction channels. The decomposition has been done according to the numbers of Z=2 particles and fragments (Z$\geq$3) in each event. The results are presented in Fig.~\ref{contrib} (upper left panel). It can be seen that the heaviest fragments (down to charge 18) are most often detected alone (no Z=2) (in fact they are accompanied by only neutrons or hydrogen isotopes) while lighter ones are generally associated with helium emission. As expected, production cross sections of IMF are mostly populated by events with at least two fragments. This explains why in Fig.~\ref{secz} this part of the spectrum is badly reproduced by classical evaporation codes as ABLA. The same decomposition has been done with the de-excitation codes mentioned above, all associated with INCL4 (three other panels of Fig.~\ref{contrib}). They are compared to the experimental data represented by the solid lines. As seen earlier, the generalized evaporation model, GEM, predicts too little IMF production. The interpretation for production cross sections of fragments with a charge between 10 and 15 is more delicate. In the experimental data, this part of the spectrum is mostly dominated by events with at least one helium particle in the final state. The fact that GEM underestimates these cross sections could be due to a lack of helium production at high $E^{\star}$, as will be shown in the following.

The case of SMM illustrates that the data obtained with SPALADIN are much more constraining than inclusive production spectra: For instance, in the region 8$\leq$Z$\leq$15, the total production cross sections are relatively well reproduced by SMM. Actually, this results from an overprediction of events with 3 fragments compensated to some extent by an underestimation of events with helium emission. Finally, we can see that the best agreement is obtained with GEMINI that reproduces rather well the different contributions although events without Z$=$2 are a little overestimated.

\begin{figure}
\begin{minipage}[c]{.48\linewidth}
\begin{center}
\includegraphics[height=4cm, angle=-90]{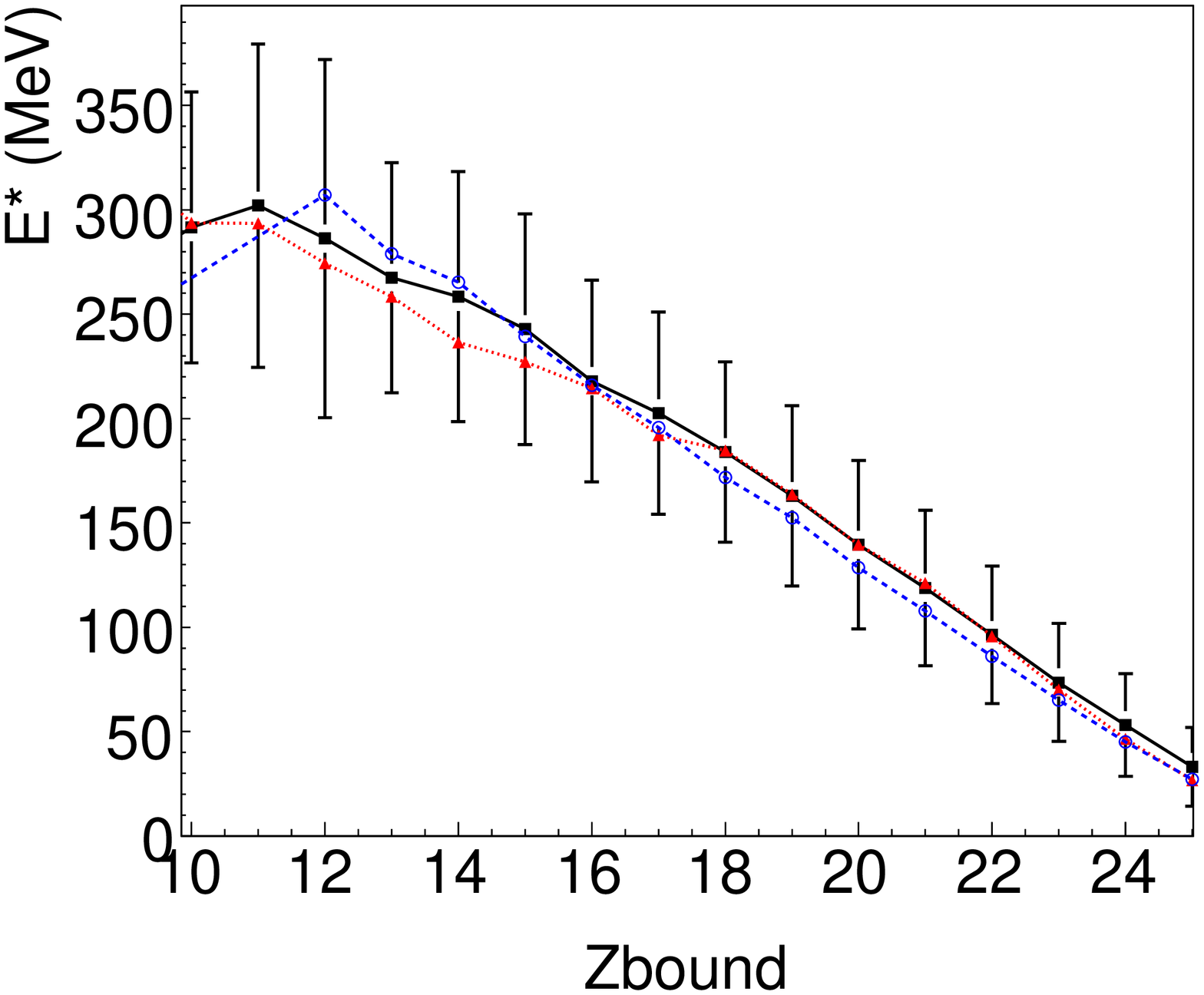}
\end{center}
\end{minipage}\hfill
\begin{minipage}[c]{.48\linewidth}
\begin{center}
\includegraphics[width=4cm]{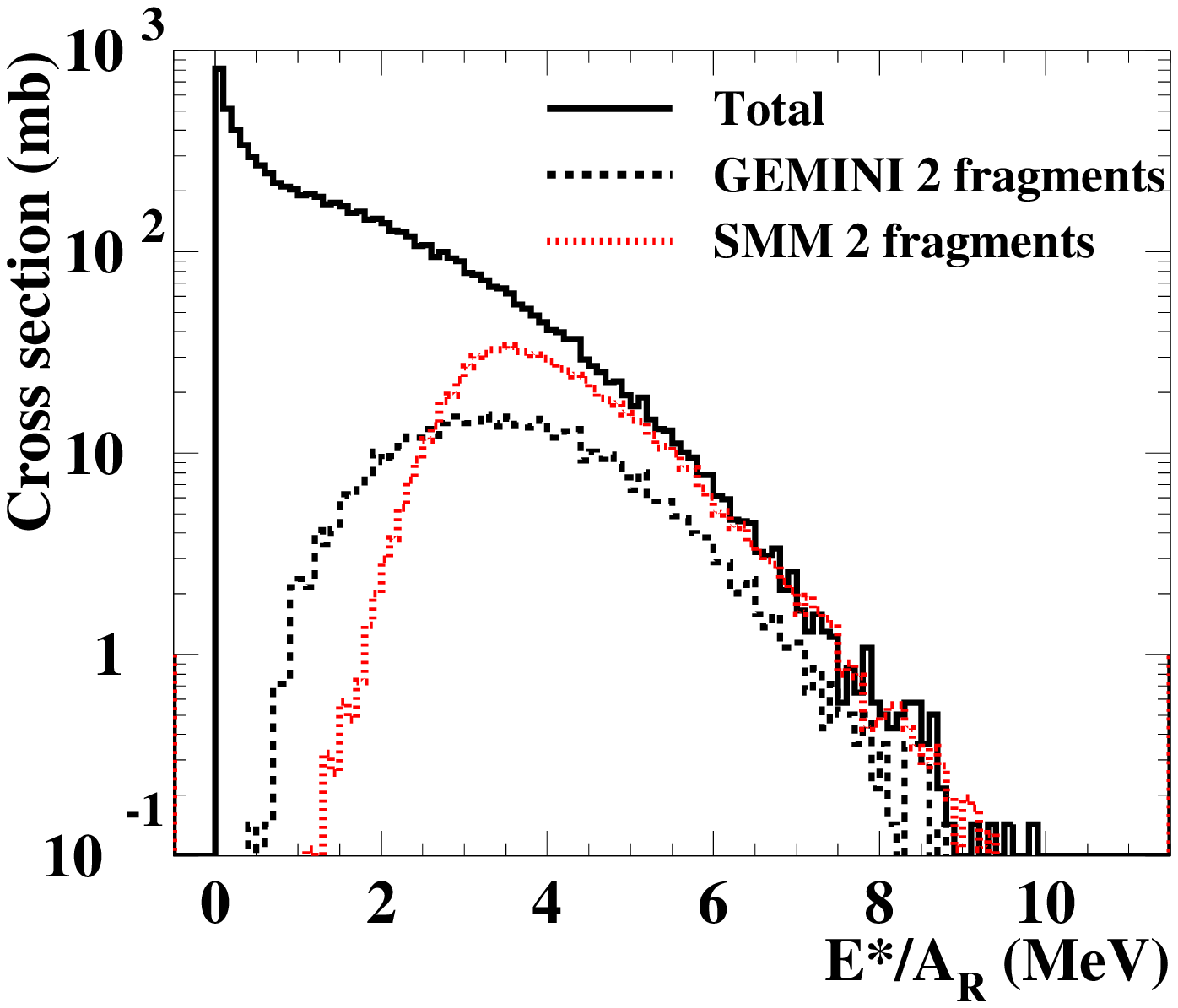}
\end{center}
\end{minipage}
\caption{\label{estarzbound}
(Color online) Left: $E^{\star}$ at the end of the cascade versus the variable $Z_{bound}$ simulated for the different de-excitation models, after GEANT4 filtering. The vertical bars indicate the standard deviation of the distributions in the case of GEMINI. Right: Distribution of $E^{\star}/A_R$ for all events (full line) and for events with at least 2 fragments for GEMINI (dashed line) and SMM (dotted line).}
\end{figure}

In order to understand the transition from classical evaporation to IMF emission, we have looked for variables that could be related to the excitation energy at the end of the cascade stage, $E^{\star}$. Two different and nearly complementary variables have been chosen: The multiplicity of neutrons plus Z=2 particles, as we mostly detect de-excitation particles (and poorly Z=1), and the variable $Z_{bound}$. The latter is defined as the sum of all charges strictly larger than one, as originally introduced by the ALADIN collaboration~\cite{HUB91}, and is related to the charge remaining bound into fragments. The choice of one or the other variable is dictated by the necessity to avoid possible correlation effects when looking at certain observables as a function of one variable. Their correlation with $E^{\star}$ has been studied with the different models filtered by GEANT4 and found to be practically independent of the de-excitation model. The case of $Z_{bound}$ can be seen on Fig.~\ref{estarzbound} left. This indicates that the range of $E^{\star}$ associated with a given bin of $Z_{bound}$, for instance, can be determined rather unambiguously. Of course, the distribution of $E^{\star}$ over all events could depend on the INC stage. The distribution of $E^{\star}$ divided by the mass of the remnant $A_R$ is displayed in Fig.~\ref{estarzbound} right. Events with $E^{\star}/A_R>4$ MeV (often claimed to be the threshold for multifragmentation~\cite{VIO06}) represents 6.5\% of the cross-section. Events with at least 2 fragments are also shown: with SMM (dotted curve), as expected for a multifragmentation model, they are associated with the highest $E^{\star}/A_R$, beginning around 3 MeV, while with GEMINI (dashed curve) they are more broadly distributed.

\begin{figure}
\includegraphics[width=8cm]{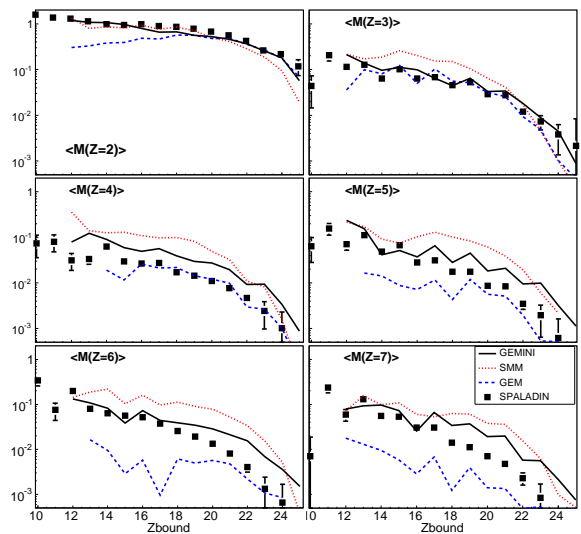}
\caption{\label{mult} (Color online) Mean multiplicities as a function of $Z_{bound}$ (explanations in text). Lines for the codes are stopped when the number of events for a given $Z_{bound}$ becomes lower than 30.}
\end{figure}

The mean multiplicities per event of the different types of fragments are shown as a function of $Z_{bound}$ on Fig.~\ref{mult}. It reveals that mean multiplicities of $Z>2$ fragments remain low, even at high $E^{\star}$. Actually, events with 2 (resp. 3) fragments represent 32$\pm$3 (resp. 2$\pm$0.2)~mb out of a total cross section of 777$\pm$79~mb. With GEMINI (resp. SMM) the cross-section of events with at least 2 fragments is 50 (resp. 80) mb. An apparent odd-even effect is observed in the mean multiplicity of $Z=3$ and $Z=5$. This is linked to the mainly binary character of the breakup and to the fact that it is unlikely to produce two fragments with both an odd number of protons.

The generalized evaporation of both GEM and GEMINI, respectively through the Weisskopf and Hauser-Feshbach formalism, reproduces very well the mean multiplicities for $Z=3$. For $Z=4$, GEM is still very good but GEMINI less. For heavier fragments, however, the multiplicities predicted by GEM remain much too low, whereas they are well reproduced by GEMINI. The comparison with SMM shows that it globally strongly overpredicts the IMF emission. As shown in \cite{VIL06}, the contribution of multifragmentation in SMM is about 30\% in the region $Z<10$, however the use of a breakup volume equal to 3 times the pre-fragment volume (code default value) is questionable in the case of spallation. In fact, the IMF multiplicities remain globally the same when the breakup volume is reduced to the pre-fragment volume. In that case, the lower number of IMF emitted through multifragmentation is globally compensated by an increasing number of IMF emitted through the SMM generalized evaporation module. The case of helium emission is also very interesting because it keeps increasing when the excitation energy increases. This is well predicted by SMM and GEMINI but not by GEM, which predicts a plateau for $Z_{bound}<20$. This behavior is linked to the competition between neutrons, protons, He and IMF emission at high excitation energy, which depends on the modeling of the Coulomb barrier, the density of states or the transmission coefficients in the different codes.

\begin{figure}
\includegraphics[width=9cm]{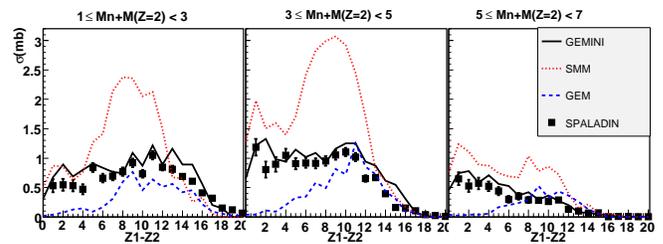}
\caption{\label{diff} (Color online) Charge difference between the 2 heaviest fragments detected in multi-fragment events, for 3 bins in the multiplicity of neutrons and Z=2 particles, corresponding to average $E^{\star}/A_R$ of respectively 3.1, 3.8 and 4.5 MeV.}
\end{figure}

In order to better understand the mechanism responsible for the IMF emission, we have looked at the difference between the highest charge, $Z1$, and the second highest charge, $Z2$, for multi-fragment events (Fig.~\ref{diff}). The events were divided into 3 bins in particle multiplicity, going from low to high excitation energies. A transition from an asymmetric breakup to a more symmetric one is observed when $E^{\star}$ is increasing. While GEM rather well reproduces the case of large $Z1-Z2$, it is totally unable to predict more symmetric breakup channels, as expected from a model with only evaporation channels. The evolution with $E^{\star}$ is better reproduced by SMM but, as mentioned before, the global IMF emission is too high and the shape at low $E^{\star}$ not correct. This could be due to the evaporation part of SMM. Only GEMINI reproduces very well the results, in shape and cross-sections. These observations indicate that evaporation, even generalized to IMF emission cannot account for the observed symmetric breakups and that a formalism like the Transition State Model could explain the production of fragments with $Z>4$ better than a model with multifragmentation, even for events corresponding to $E^{\star}/A_R \simeq$ 4 MeV. This does not support the conclusion of~\cite{NAP04} based on velocity measurements that IMF are formed in simultaneous breakup decays.

To conclude, the simultaneous measurement of light particles and spallation residues, performed in inverse kinematics with the SPALADIN setup, on the system Fe+p at 1~GeV per nucleon, has allowed for the first time the decomposition of the total reaction cross-section into the different decay channels and the study of its evolution with excitation energy.  The comparison with the predictions of different de-excitation models coupled to INCL4 confirmed that the observed high production cross-sections of IMF, already observed in inclusive measurements, cannot be explained by standard evaporation models, even generalized to the emission of IMF, as GEM. This model does not predict the appearance of symmetric breakups observed when the excitation energy increases. The use of models with other mechanisms for the IMF production, such as multifragmentation in SMM or the Transition State Model asymmetric splitting modes of GEMINI, helps to reproduce the data. It appears that the best description of the whole set of coincidence data is obtained with the INCL4+GEMINI code, in particular as regards the dependence with excitation energy. It seems therefore that in a light system as Fe+p, with no compression and little angular momentum ($9\hbar$ in average) multifragmentation might not be necessary to explain the data. The better understanding of the reaction mechanism reached through coincidence measurements of the present kind will certainly allow the development of more reliable simulation tools for applications of spallation reactions.

This work was supported by the EU under contract No. HPRI-CT-1999-00001.

\bibliography{LeGentil_bib}

\end{document}